# High Speed, Low Power Current Comparators with Hysteresis

Neeraj K. Chasta

Dhirubhai Ambani Institute of Iinformation & Communication Technology, Gandhinagar, Gujarat

neeraj.chasta@gmail.com

## ABSTRACT

*This paper, presents a novel idea for analog current comparison which compares input signal current and reference currents with high speed, low power and well controlled hysteresis. Proposed circuit is based on current mirror and voltage latching techniques which produces rail to rail output voltage as a result of current comparison. The same design can be extended to a simple current comparator without hysteresis (or very less hysteresis), where comparator gives high accuracy (less than 50nA) and speed at the cost of moderate power consumption. The comparators are designed optimally and studied at 180nm CMOS process technology for a supply voltage of 3V.*

## 1. INTRODUCTION

Nodal voltages and branch currents represents the information carried by any electric network. The former referred to "voltage domain circuits" whereas the latter are known as "current domain circuits". In recent years; as supply voltage and device threshold voltage reduction greatly affected the performance of voltage domain circuits, current mode circuits are becoming designer's interest. Current domain circuits show many unique and attractive properties over their voltage mode complements including higher speed, higher bandwidth, reduced distortion, low supply voltage requirements and lesser sensitivity to switching noise.

Several current mode comparison approaches are proposed [5]-[14]. The first CMOS continuous time current comparator was proposed in [5]. It consists of two cascode current mirrors but can't operate at high speeds and frequencies due to high output impedance. To overcome this problem a new circuit has been proposed [7] which uses inverter stage in feedback with source-follower stage. But the circuit shows dead band region for low values of currents, where input impedance is quite high and thus limits speed of operation.

The input and output impedances are further controlled by voltage-current feedback concept [12], [14]; where a resistive feedback is applied to a voltage amplifier, thus reducing impedances and improving speed performance. Current mirror concept again used in [15]; where it uses improved Wilson current mirror for current comparison, but circuit suffers from higher power consumption and delay introduced by gain circuitry.

DOI : 10.5121/vlsic.2012.3107 85

International Journal of VLSI design & Communication Systems (VLSICS) Vol.3, No.1, February 2012

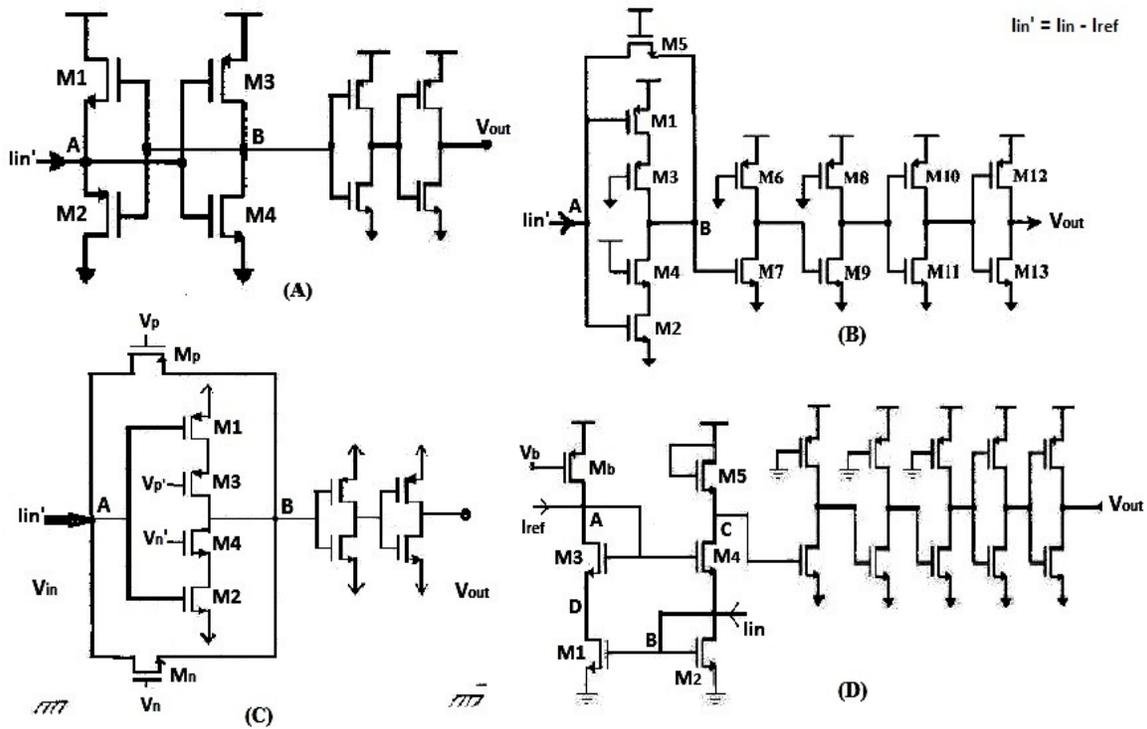

Fig. 1: Existing Current Comparators (A) H. Traff's[7] (B) R. Del's[10] (C) L. Chen's[14] (D) V. Kasemsuwan's[15]

## 2. COMPARATORS WITH HYSTERESIS

Z. Wang proposes current comparator with hysteresis [6] shown in fig 2. It uses three current sources namely "$I_1$", "$I_2$" and "$I_{hy}$". $I_1$ and $I_2$ are input currents to be compared whereas amount of hysteresis generated decided by current $I_{hy}$.

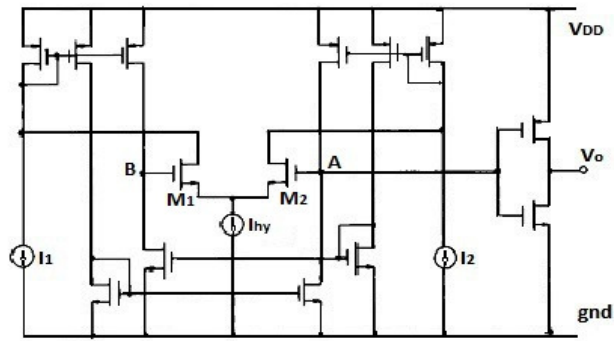

Fig. 2: Z. Wang's Comparator (with hysteresis) [6]

This circuit suffers from some major operational problems such as; functionality depends on the digital levels generated at node "A" and node "B", and does not provide any assistance at the times of transitions as both of transistors $M_1$ and $M_2$ will be in "on" state. Speed performance is low; also design needs modification for zero and negative values of currents.

## 3. CIRCUIT DESCRIPTION

Here a new comparison scheme is proposed which consists of three stages: the input preamplifier, a positive feedback latch stage and an output inverter. The preamplifier stage allow the input currents to flow through it and develop corresponding voltage at input nodes which results in terms of current output by current mirroring concept. It also isolates the input of comparator from switching noise ("kickback" noise) coming from positive feedback stage.

The latch stage determines which of the input signal is larger by positive feedback action. It is also called as "decision making stage". The inverter stage amplifies the information provided by decision stage and outputs a CMOS compatible voltage signal (0 or $V_{DD}$).

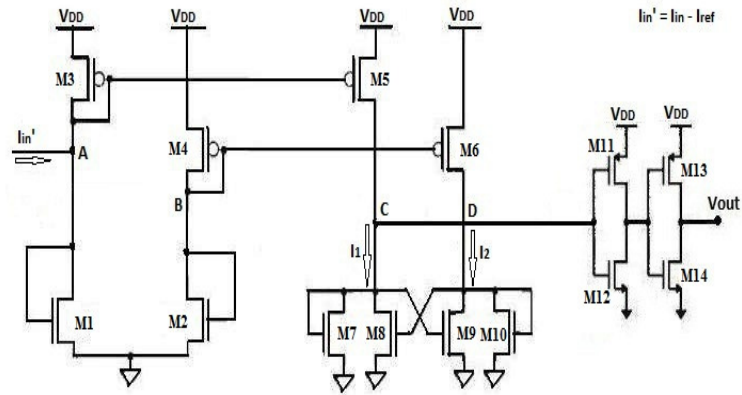

Fig. 3: Proposed Current Comparator

**Working:** Consider the case when $I_{ref}$ is constant and $I_{in}$ change from "low" to "high". A constant $I_{ref}$ will keep node B at constant voltage and allow $M_6$ to supply a constant drain current ($I_2$) to latch circuit. On the other hand change in $I_{in}$ will make node "A" voltage varying and so the gate voltage of $M_5$.

As $I_{in}$ increases, $V_A$ increases and thus reduces the drain current ($I_1$). This decrease in $I_1$ will reduce current through $M_7$ and so the voltage at node "C" ($V_C$). This change will affect the current through $M_9$ and as $I_2$ is constant, current through $M_{10}$ will increase. As drain current of $M_{10}$ is increasing, ($V_{gs}$) of $M_{10}$ will increase and so voltage at node "D" ($V_D$), positive feedback action gives produce more reduction in drain current of $M_7$ and hence in $V_C$. Variations at node "C" are applied to inverter circuitry which results in CMOS compatible rail to rail output voltage.

The circuit uses positive feedback from the cross-gate connection of $M_8$ and $M_9$ to increase the gain of latch circuitry (As circuit is symmetric $g_{m7} = g_{m10}$ & $g_{m8} = g_{m9}$). But if $g_m$'s of all the latch transistors are not equal, will lead circuit to show hysteresis property. This we can easily analyze by small signal analysis of latch circuitry.

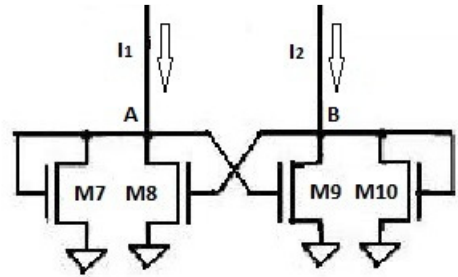

Fig. 4: Positive Feedback Latch Circuit [2]

The expressions for node voltage "A" and "B" are as (where $g_{m7} = g_{m10}$ & $g_{m8} = g_{m9}$):

$$V_A = \frac{g_{m7}}{g_{m7}^2 - g_{m9}^2}\left[I_1 - \frac{g_{m9}}{g_{m7}}I_2\right]$$

And,

$$V_B = \frac{-g_{m9}}{g_{m7}^2 - g_{m9}^2}\left[I_1 - \frac{g_{m7}}{g_{m9}}I_2\right]$$

Both the equation shows that circuits will lead to hysteresis property (if $g_m$'s of transistors $M_7$ and $M_9$ are not equal). The large signal (DC) analysis expressions for node voltage at "A" and "B" would be as:

$$V_A = V_{tn} + \sqrt{\frac{K_{n7}}{K_{n7}^2 - K_{n9}^2}\left(I_1 - \frac{K_{n9}}{K_{n7}}I_2\right)}$$

And,

$$V_B = V_{tn} + \sqrt{\frac{K_{n9}}{K_{n7}^2 - K_{n9}^2}\left(\frac{K_{n7}}{K_{n9}}I_2 - I_1\right)}$$

Here we have expressions for $V_A$ and $V_B$ in large and small signal analysis, and can see that in positive feedback configuration ($g_{m7} < g_{m9}$) circuit will show hysteresis. From the discussion above we can see that for proposed design:

$$(V_C - V_{th})^2 = -\frac{K_{n7}K_{p5}}{K_{p3}(K_{n7}^2 - K_{n9}^2)}\left[I_{in} - I_{ref} + \frac{K_{n9}}{K_{n7}}I_{D2} - I_{D1}\right] \quad ..(A)$$

And,

$$(V_D - V_{th})^2 = \frac{K_{n9}K_{p5}}{K_{p3}(K_{n7}^2 - K_{n9}^2)}\left[I_{in} - I_{ref} + \frac{K_{n7}}{K_{n9}}I_{D2} - I_{D1}\right] \quad ..(B)$$

Here both the equations show that proposed circuit will produce hysteresis in output voltage with respect to applied input currents and hysteresis produced depends on drain current of $M_2$ ($I_{D2}$). The exact calculation of hysteresis generated is explained in next section.

## 4. CALCULATION OF HYSTERESIS

As we are using positive feedback latch circuit, at the time transitions one of the cross coupled transistor will go in triode region and so the amount of hysteresis introduced will also depend on the drain voltages of $M_8$ and $M_9$ ($V_C$ and $V_D$ respectively).

When current changes from "low" to "high" values at the time of switching transistor "$M_8$" will be in triode keeping all other transistors in saturation. So the ratio of drain currents of $M_5$ and $M_6$ can be given by:

$$\frac{I_{D5}}{I_{D6}} = \frac{(V_C - V_{tn})^2 + (K_9/K_7)(2V_D - 2V_{tn} - V_C)V_C}{(V_D - V_{tn})^2 + (K_9/K_7)(V_C - V_{tn})^2} = P$$

As "$I_{D5}$" and "$I_{D6}$" are mirrored by $M_3$ and $M_4$ respectively, so

$$\frac{I_{D5}}{I_{D6}} = \frac{I_{D3}}{I_{D4}} = P$$

Appling KCL at input nodes A and B, we can easily see that

$$I_{in} = I_{ref} + I_{D1} - PI_{D2}$$

Input current in equation (3.36), give the value of current at the time of transition and it can be referred as transition current ($I_{t1}$).

Same can be done for reversed case; when current changes from "high" to "low" values, this time transistor "$M_9$" will go in triode region and input signal current at the time of switching can be given by ($I_{t2}$):

$$I_{in} = I_{ref} + I_{D1} - P'I_{D2}$$

Where "$P'$" can be given by,

$$P' = \frac{(V_C - V_{tn})^2 + (K_9/K_7)(V_D - V_{tn})^2}{(V_D - V_{tn})^2 + (K_9/K_7)(2V_C - 2V_{tn} - V_D)V_D}$$

So we have transition currents expressions as,

$$I_{t1} = I_{ref} + I_{D1} - PI_{D2} = I_{ref} + I_a$$

$$I_{t2} = I_{ref} - (P'I_{D2} - I_{D1}) = I_{ref} - I_b$$

Equations derived above can be implemented graphically in hysteresis curve as:

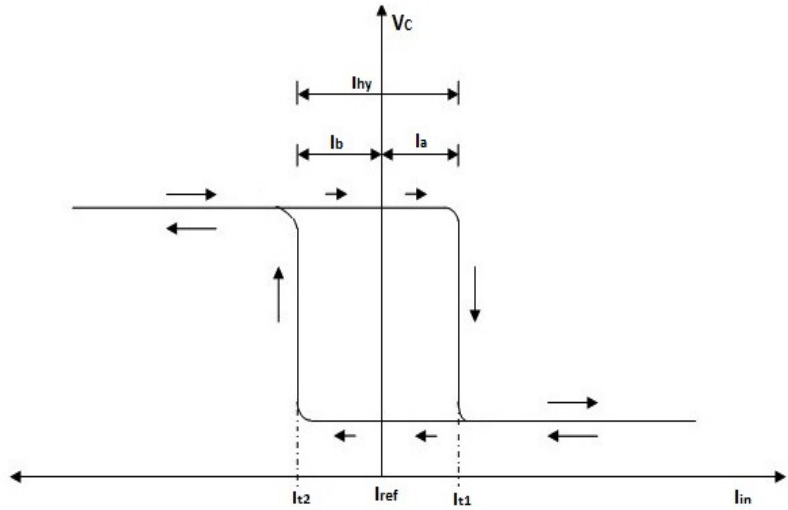

Fig. 5: Hysteresis Curve ($V_C$ v/s $I_{in}$)

Then hysteresis current can be given by

$$I_{hy} = |I_{t1} - I_{t2}|$$

$$I_{hy} = |P'I_{D2} - PI_{D2}|$$

Equation shows the expression of hysteresis produced by circuit; where we see that it is well controlled as drain current of $M_2$ ($I_{D2}$) is always constant (see fig.3 $M_2$ and $M_4$ are diode connected and no current changing element is available).

Keeping equations (A) and (B) in mind; we see that when all the transistors are in saturation, amount of hysteresis produced depends on the Ration of "Kn's". If we set this ratio near to 1; so that $\left[\left(K_{n9}/K_{n7}\right)I_{D2}\right]$ term approximately get cancelled with "$I_{D1}$", and thus amount of hysteresis introduced can nearly be set to zero. This we can understand in other way around that if we design this circuit in such a way that gm's of all the latch transistors be nearly equal ($g_{m7} \cong g_{m9}$ or $K_{n7} \cong K_{n9}$) hysteresis produced can be reduced to "zero". As we have already discussed that hysteresis introduced directly relates with depth of positive feedback (difference of $g_m$'s), then this will be the condition when circuit will have a slight positive feedback and so very less hysteresis. Under such stipulations circuit can also be used as a simple current comparator.

## 5. SIMULATION RESULTS

For simulation, standard BSIM 0.18µm CMOS technology parameters have been used with 3V power supply. Proposed circuit was designed optimally for speed, power, accuracy and hysteresis.

**Comparator with hysteresis**: The design proposed produces hysteresis in output voltage with respect to the change in direction of input current. DC transfer characteristics observed with increasing and decreasing values of input currents are shown in Fig. 6 & 7 respectively.

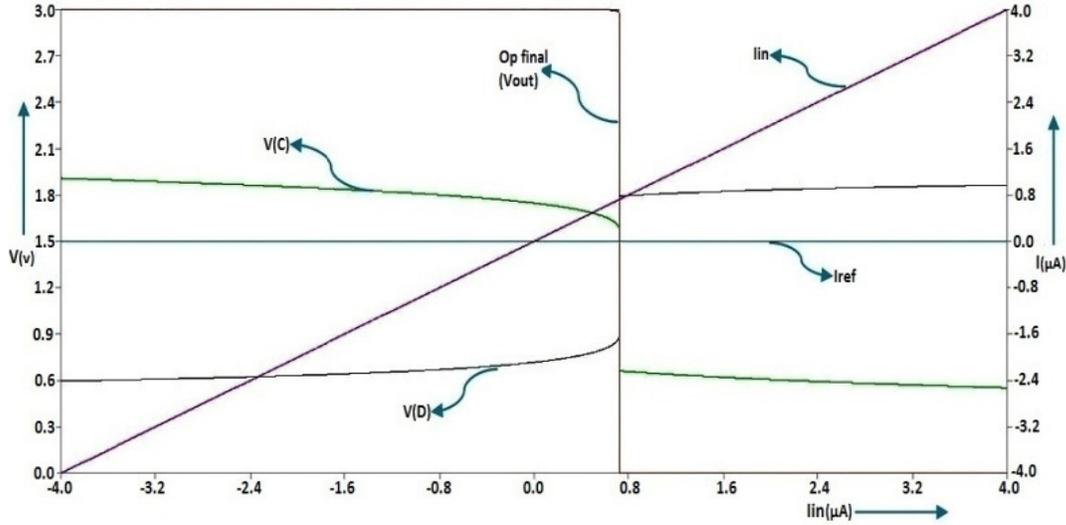

Fig. 6: Hysteresis for Increasing value of $I_{in}$ ($I_{ref} = 0\mu A$)

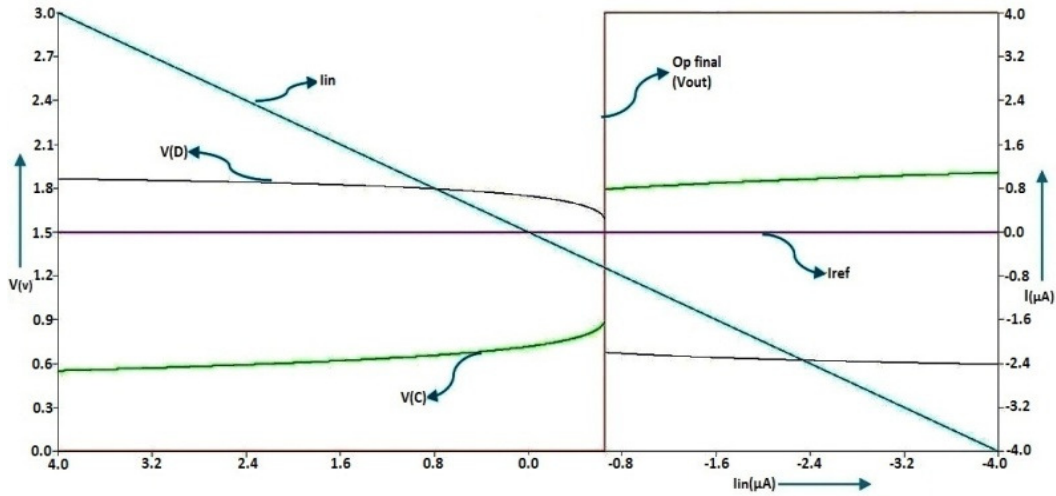

Fig. 7: Hysteresis for Decreasing value of $I_{in}$ ($I_{ref} = 0\mu A$)

Here in figures, we see that switching of output occurs at 0.72µA (Fig. 6) and -0.65µA (Fig. 7) depending on the direction of change in current. So total amount of hysteresis produced:

$$I_{hy} = |-0.7\mu A| + 0.7\mu A = 1.4\mu A$$

Circuit produces average propagation delay of 0.2ns and 4.8ns for +/- 100µA and +/- 1µA input square-wave current pulses respectively, at the DC power consumption of 1.8mW. Transistor dimensions used are as:

Table 1: Design Parameters: Comparator with Hysteresis

|  | W (Width) | L (Length) |
|---|---|---|
| **M1, M2** | 0.18µm | 0.72µm |
| **M3, M4** | 0.54µm | 0.72µm |
| **M5, M6** | 1.08µm | 0.18µm |
| **M8, M9** | 0.36µm | 0.18µm |
| **M7, M10** | 0.27µm | 0.18µm |
| **Inverter: PMOS** | 0.54µm | 0.18µm |
| **Inverter: NMOS** | 0.18µm | 0.18µm |

**Comparator without hysteresis**: The same design could be extended to comparator without (or very less) hysteresis in output voltage with the change in direction of input current. DC transfer characteristics obtained for extended concept with increasing and decreasing values of input currents are shown in Fig. 8 & 9 respectively.

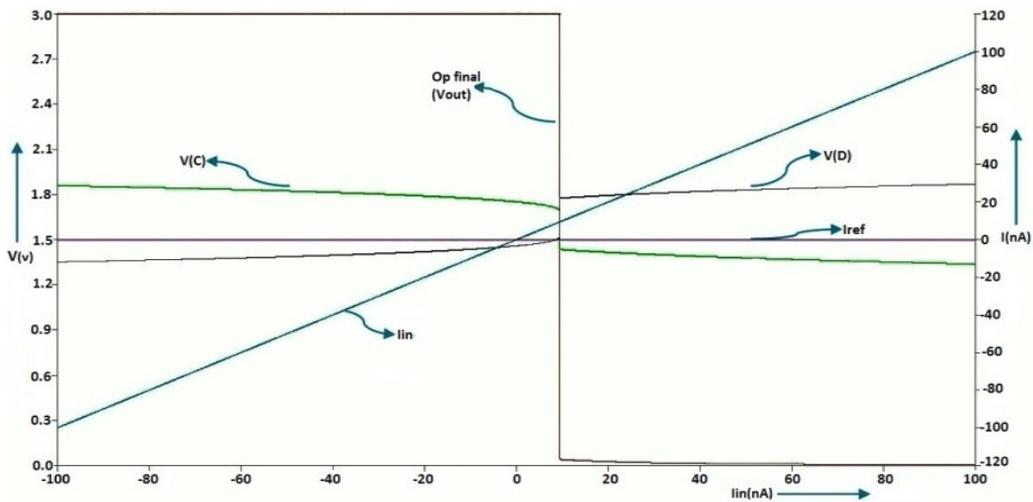

Fig. 8: Hysteresis for Increasing value of $I_{in}$ ($I_{ref}$ = 0nA)

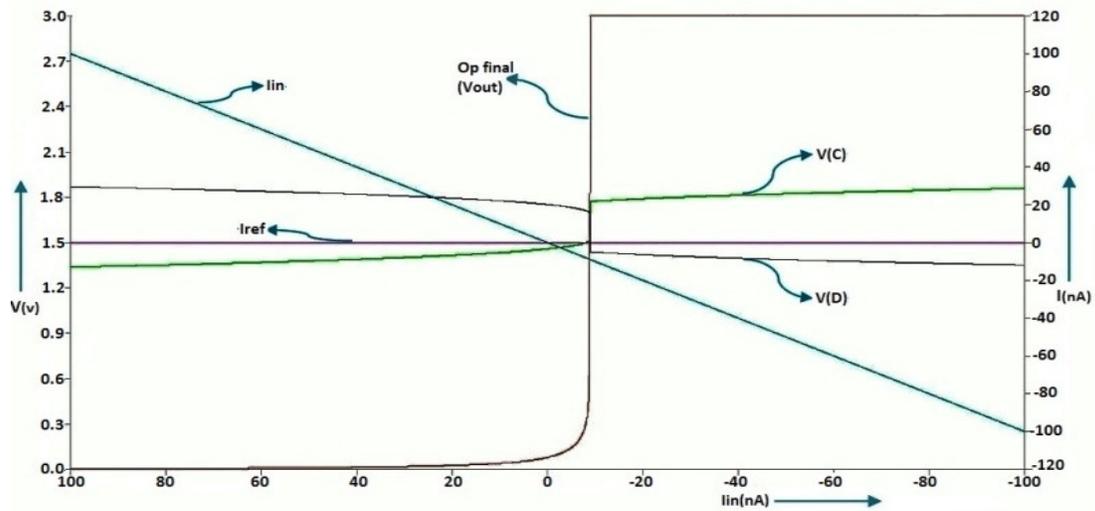

Fig. 9: Hysteresis for Decreasing value of $I_{in}$ ($I_{ref}$ = 0nA)

Here, we see that for this design switching of output doesn't much affected by the direction of change in current as the transition points are at 9nA (Fig. 8) and -9nA (Fig. 9) with rising and falling values of input current, respectively. This is just because of very less positive feedback. Total amount of hysteresis produced:

$$I_{hy} = |-9nA| + 9nA = 18nA$$

Circuit produces average propagation delay of 0.2ns and 1.6ns for +/- 100µA and +/- 1µA input square-wave current pulses respectively, at the DC power consumption of 2.3mW. Transistor dimensions used are as:

Table 2: Design Parameters: Comparator without Hysteresis

|  | W (Width) | L (Length) |
|---|---|---|
| **M1, M2** | 0.18µm | 0.72µm |
| **M3, M4** | 0.18µm | 0.72µm |
| **M5, M6** | 1.19µm | 0.18µm |
| **M8, M9** | 0.34µm | 0.18µm |
| **M7, M10** | 0.21µm | 0.18µm |
| **Inverter: PMOS** | 0.54µm | 0.18µm |
| **Inverter: NMOS** | 0.18µm | 0.18µm |

To compare the performance of proposed comparator (without hysteresis case) with existing low power comparators i.e. Traff's [7], R. Del's [10], L. Chen's [14] and V. Kasemsuwan's [15], LTSPICE simulations of all comparators are performed using standard BSIM 0.18µm CMOS technology parameters with 3V power supply.

93

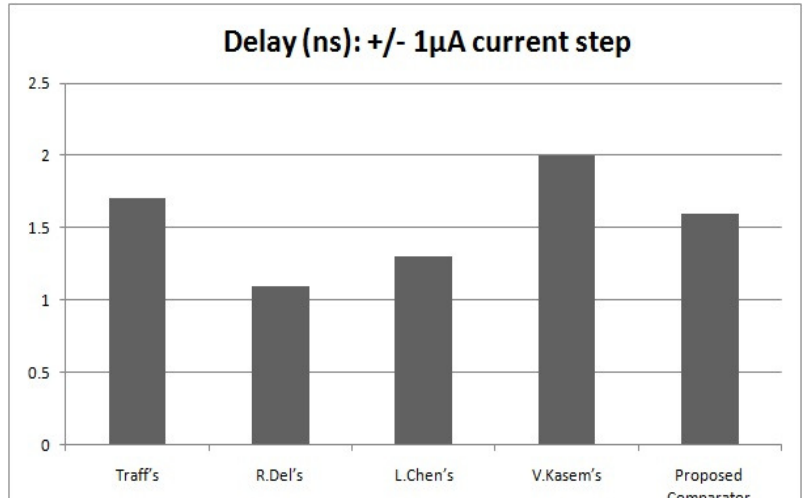

Fig. 10: Average Delay (ns): Response to +/- 1µA Current Step

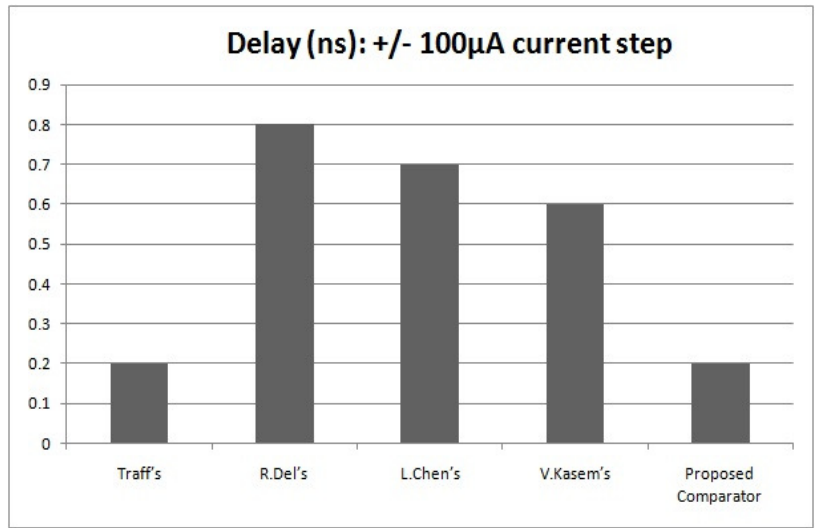

Fig. 11: Average Delay (ns): Response to +/- 100µA Current Step

## 6. CONCLUSION

A new current mode comparator with hysteresis is presented. Design have novel structure and able to achieve crucial parameter of hysteresis control. It is having analogous functionality and far better performance over the existing one in terms of speed and hysteresis control.

The same design can be extended to a simple current comparator, which compares input signal and reference currents with high accuracy (less then 50nA), high speed and comparable power consumption with all existing low power comparators.

## ACKNOWLEDGEMENT

I sincerely thank Prof. Chetan Parikh for his valuable guidance in the study of Comparators.

**Authors**

Neeraj K. Chasta received his B. Eng. Degree i n Electronics & Communication from University of Rajasthan, Jaipur, India in 2007, and the M. Tech degree in Information & Communication Technology with specialization VLSI from Dhirubhai Ambani institute of Information & Communication Technology (DA-IICT), Gandhinagar, India in 2010. His current research interests are in analog & mixed signal design with low power & high speed emphasis & VLSI Testing.

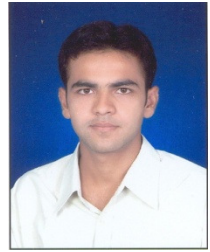